\documentclass[prd,nofootinbib,twocolumn]{revtex4-1}
\usepackage{amsmath}
\usepackage{amssymb}
\usepackage[bookmarksopen,bookmarksnumbered]{hyperref}

\let\oldleft\left
\def\xleft{\mathopen{}\oldleft}

\newcommand\3[1]{\boldsymbol{#1}}

\newcommand{\MSbar}{\ensuremath{ \overline{\rm MS} }}

\newcommand\BIBvol[1]{\textbf{#1}}

\begin{document}


\title{CSS Equation, etc, Follow from Structure of TMD Factorization}

\author{John Collins}
\email{collins@phys.psu.edu}
\affiliation{%
  104 Davey Lab, Penn State University, University Park PA 16802, USA
}


\date{26 November 2019}

\begin{abstract}
  I show that the forms of the Collins-Soper-Sterman and
  renormalization-group equations for the evolution of
  transverse-momentum-dependent (TMD) parton densities in QCD follow
  from the structure of TMD factorization.  A derivation does not need
  to directly use detailed properties of the operator definition of
  the TMD parton densities.
\end{abstract}

\maketitle


\section{Introduction}
\label{sec:intro}

An accurate description of the Drell-Yan process and other similar
processes requires an understanding in QCD of the $Q$ dependence of
the transverse-momentum dependence when $q_T \ll Q$.  (Here, $q_T$ and
$Q$ are the transverse momentum and invariant mass of the Drell-Yan
pair.)  An often-used way of treating this issue was found by Collins
and Soper (CS) \cite{Collins:1981uw, Collins:1981uk}, and Collins,
Soper and Sterman (CSS) \cite{Collins:1984kg}, who used factorization
with transverse-momentum-dependent parton density functions (TMD
pdfs).  In contrast to ordinary ``integrated pdfs'', the TMD pdfs of
CS depend on an extra argument $\zeta$.  The phenomenological significance
of $\zeta$ can seen in the basic factorization formula, Eq.\
(\ref{eq:TMD.fact}) below, where the two instances of $\zeta$ obey
$\zeta_A\zeta_B=Q^4$.  CS obtained an equation for the $\zeta$ dependence.  This
equation implies a strong $Q$-dependence of the shape of the measured
$q_T$ distribution, which is in agreement with data for the Drell-Yan
and other processes --- see, for example, Refs.\ \cite{Landry:2002ix,
  Konychev:2005iy, Aybat:2011ta, Anselmino:2012aa}.

However, there is a bewildering variety of treatments of TMD
factorization in the literature, with what for the outsider appear to
be contradictory claims about the need for and the use of the CS
equation and other aspects of the CSS approach.  This can be seen in
work within an SCET framework by Becher and Neubert (BN)
\cite{Becher:2010tm} and by Echevarr\'{\i}a, Idilbi, and Scimemi (EIS)
\cite{GarciaEchevarria:2011rb}.  Notably, the last authors state that
their approach is ``without the Collins-Soper like evolution
equations''.  In reality, however, the last authors do use an exact
equivalent of the CS equation, but label it differently.  To see this,
one can merely compare Eq.\ (5.7) of Ref.\
\cite{GarciaEchevarria:2011rb} with the results in Sec.\ 13.13.1 of
\cite{Collins:2011qcdbook}, which uses a modernized and improved
version of the CSS method.  One difference in the EIS treatment is the
derivation of the results, which are shown to follow from more
structural properties of TMD factorization, but within the particular
approach of EIS.  The strong $Q$-dependence of the transverse-momentum
distribution is a property of QCD that cannot be avoided.  The most
that can be done is to change the method of derivation and the form of
presentation of the results.  (Another set of discrepancies between
the treatments concerns the large-$b_T$ region, where $b_T$ is a
variable Fourier-conjugate to transverse momentum.  In this region,
non-perturbative information is needed beyond that in the integrated
pdfs.  A discussion of the disagreements on this topic is not the
concern of this paper.)

In this paper, motivated by a derivation by EIS, I show that the form
of the CS equation and associated renormalization group (RG)
properties follow from structural properties of TMD factorization (in
its recent formulations).  In contrast, the derivations by CS
\cite{Collins:1981uw,Collins:1981uk} and by Collins
\cite{Collins:2011qcdbook} used specific properties of the operators
defining the TMD pdfs to obtain their dependence on the $\zeta$ parameter,
and therefore appear to tie the CS equations to these properties of
the operators.  Thus these derivations are inapplicable, for example,
to the rather different definition given by EIS
\cite{GarciaEchevarria:2011rb}, even though the definitions give
results numerically equivalent \cite{Collins:2012uy} to those from the
definitions in \cite{Collins:2011qcdbook}.  Of course, the validity of
factorization strongly constrains which operator definitions of pdfs
are suitable.

The results in the present paper are important step in unifying the
various treatments of TMD factorization, as regards their
phenomenological implications, following on from
\cite{Collins:2012uy}.  Note that the derivations by BN
\cite{Becher:2010tm} and by EIS \cite{GarciaEchevarria:2011rb} are
explicitly restricted to $\Lambda_{\rm QCD} \ll q_T \ll Q$, i.e., they exclude
the region of the smallest $q_T$ where the cross section is largest.
In contrast the derivation of the form of the CS equation in this
paper applies generally, and the TMD factorization property used is
valid over the whole range of small transverse momentum $q_T \ll
Q$.\footnote{As explained, for example, in the CS and CSS work, TMD
  factorization can be matched with ordinary collinear factorization
  at large $q_T$ to give results valid for all $q_T$.}  Note also that
BN \cite{Becher:2010tm} did not give definitions of finite TMD pdfs
within their approach: They obtained finite results only for the
product of two TMD functions.

Among other results that are associated with the CS equation and its
derivation are that the anomalous dimension of the TMD pdf has linear
dependence on $\ln(\zeta/\mu^2)$ (or equivalently $\ln (Q^2/\mu^2)$), where
$\mu$ is the RG scale.  This logarithmic dependence is controlled by the
$\zeta$-independent RG coefficient $\gamma_K$ for the kernel of the CS
equation.  Another result is that the kernel of the CS equation does
not depend on $\zeta$ (or $Q$) or on the flavor of the partons and beam
hadrons.  Both of these results also follow from the derivation in
this paper, which does not in any way invalidate the derivations in
\cite{Collins:1981uw, Collins:1981uk, Collins:2011qcdbook}.

\section{What the CS and Collins derivations did}

CS \cite{Collins:1981uw, Collins:1981uk} defined TMD pdfs by matrix
elements of appropriate non-gauge-invariant operators in a
\emph{non}-light-like axial gauge with gauge fixing vector $n$.  This
vector can be thought of as providing a standard for classifying
momenta into left-movers and right-movers.  They found it convenient
to use the 
variable $\zeta=4(xP\cdot n)^2/|n^2|$ for parameterizing the dependence on the
vector $n$, where $x$ is the normal longitudinal momentum fraction in
a pdf and $P$ is the momentum of the target hadron.

Dependence on $\zeta$ is equivalent to dependence on the gauge-fixing
vector, and the CS equation was derived by using Ward identities
associated with gauge-dependence to get a particularly simple form.
The actual form of the CS equation was not known ahead of time, and
the derivation amounted to a calculation of the $\zeta$ dependence of a
TMD pdf to leading power at large $\zeta$.  The form of the CS equation,
and the various auxiliary properties, were outputs of the derivation.

Much more recently Collins \cite{Collins:2011qcdbook} worked out an
improved gauge-invariant definition of TMD pdfs, again with an
auxiliary vector $n$ which is now the direction of a certain Wilson
line.  With this definition, the CS equation arises from factorization
properties of Wilson line operators when there are two Wilson lines of
very different rapidities.  An important quantity is the anomalous
dimension $\gamma_K$, for the kernel of the CS equation; it corresponds to
a similar quantity in the work of Korchemsky and Radyushkin
\cite{Korchemsky:1987wg} for the dependence of the anomalous dimension
of a Wilson line cusp on the hyperbolic angle of the cusp.

\section{Corresponding derivation for RG equation for operators}
\label{sec:OPE.to.anom.dim}

Before examining how TMD factorization implies the CS equation, I
illustrate the principles of the method by constructing an analogous
argument for a more well-known and simpler case, which is the RG
equation for operators that appear in the short-distance
operator-product expansion (OPE).  A simple generalization of the
argument allows one to obtain the DGLAP equation for integrated pdfs
from an ordinary collinear factorization formula.

The material in this section is not original.  I remember learning the
method from someone else many years ago, but I don't remember who, or
if and where it has been published.

Let us examine the OPE for an object of the following form:
\begin{multline}
  \label{eq:F}
  F(Q,P,H,m(\mu),\mu,\alpha_s(\mu)) 
\\
  = \int d^4x e^{iq\cdot x} \langle P, H | T j(x) j(0) | P, H \rangle ,
\end{multline}
which is the Fourier transform of a current-current matrix element in
a target state of momentum $P$ and flavor $H$.  I restrict the
situation to be simple enough that only a single operator appears in
the OPE at leading power in $Q=|q|$, and, as usual, choose the
currents to be RG invariant.  Also as usual, the renormalization scale
is $\mu$, the renormalized coupling of the theory is $\alpha_s(\mu)$, and the
renormalized mass parameter(s) of the theory are denoted by $m(\mu)$

To formulate the OPE, we take an essentially Euclidean limit where $q$
is made large.  Then, up to power-suppressed corrections:
\begin{equation}
  \label{eq:OPE}
  F = C(Q,\mu,\alpha_s) \, A(P,H,m,\mu,\alpha_s),
\end{equation}
where $C$ is a short-distance coefficient function and $A$ is the
expectation value of some renormalized operator $\mathcal{O}$ in the
state $|P\rangle$: $A = \langle P,H|\mathcal{O}|P,H\rangle$.  The operator is a product
of fields (and possibly derivatives) at some space-time point, and of
a renormalization factor.

Important properties of this statement of the OPE are that: (a) The
coefficient function depends on $Q$ and $\mu$, but not on the mass(es)
or on $P$ or on hadron flavor $H$.  (b) The operator matrix element is
independent of $Q$.  Furthermore, if the currents defining $F$ are
changed such that the same operator $\mathcal{O}$ is needed in the
OPE, then the value of $C$ can depend on which currents are used, but
the matrix element $A$ is independent of the choice of currents.  Thus
the only arguments in common between the two factors are $\mu$ and
$\alpha_s$. These properties are a consequence of deriving the OPE by a
strict expansion in powers of $Q$ and keeping only the leading power,
and of using a mass-independent renormalization scheme like \MSbar.

\subsection{Standard derivation from renormalization properties}
\label{sec:anom.dim.std}

The standard derivation of the RG equation for the operator and its
matrix element $A$ uses the fact that the operator is multiplicatively
renormalized, so that the bare operator and the renormalized operator
are related by $\mathcal{O}_0 = Z_A \mathcal{O}$.  The renormalization
factor $Z_A$ is a function of $\alpha_s$ and the dimensionless regulator
parameter $\epsilon$.  Then the renormalization of the matrix element
similarly obeys
\begin{equation}
  A_0 = Z_A A.
\end{equation}
Since bare operators are RG invariant, the RG equation for the
renormalized matrix element is obtained by
\begin{equation}
  \label{eq:RG.A}
  \frac{ dA }{ d\ln\mu } = 
  \frac{ dZ_A^{-1} }{ d\ln\mu } A_0
  = -\gamma_A(\alpha_s) A,
\end{equation}
where
\begin{equation}
  \gamma_A(\alpha_s) = \frac{ d\ln Z_A }{ d\ln\mu }.
\end{equation}
Here a total derivative is used, as usual, to indicate that the
differentiation with respect to $\mu$ includes the running of the
coupling and of the quark masses:
\begin{equation}
    \frac{ d }{ d\ln \mu }
    = \frac{ \partial }{ \partial\ln\mu }
      + \frac{ d\alpha_s(\mu) }{ d\ln \mu } \frac{ \partial }{ \partial\alpha_s }
      + \sum_i \frac{ dm_i(\mu) }{ d\ln \mu } \frac{ \partial }{ \partial m_i }.
\end{equation}
The anomalous dimension $\gamma_A$ is a function of the renormalized
coupling alone.

Since the current-current matrix element $F$ is RG invariant, so is
the leading term $CA$ in its expansion in powers of $Q$.  Hence the RG
equation of the coefficient $C$ has the opposite anomalous dimension
to $A$:
\begin{equation}
  \label{eq:RG.C}
  \frac{ dC }{ d\ln\mu } 
  = \gamma_A(\alpha_s) C.
\end{equation}

For applications, the important result is that the anomalous dimension
depends only on $\alpha_s$.  Thus the condition that a perturbative
calculation of $\gamma_A$ is accurate is simply that $\alpha_s$ is small enough;
there are no large logarithms of kinematic variables to worsen the
perturbation expansion, unlike the coefficient function $C$ with its
logarithms of $Q/\mu$.  We apply the RG equation to express the product
$CA$ in terms of $C$, with $\mu$ set equal to $Q$, and of $A$ with a
fixed reference value $\mu_0$ of the renormalization scale:
\begin{multline}
\label{eq:RG.solution}
  C \, A = C(Q,Q,\alpha_s(Q)) \, A(P,H,m,\mu_0,\alpha_s)
\times \\ \times
    \exp\left\{ - \int_{\mu_0}^Q \frac{d\mu}{\mu} \gamma_A(\alpha_s(\mu)) \right\} .
\end{multline}
On the right-hand-side, $C$ has no large logarithms, so that it may be
calculated perturbatively if $\alpha_s(Q)$ is small, i.e., if $Q$ is large
enough, because of the asymptotic freedom of QCD.  Since $A$ is at a
fixed renormalization scale, the formula illustrates the universality
properties of non-perturbative part embodied in the operator matrix
element.

All the above is (or should be) well-known and can be found in
textbooks.

\subsection{Derivation from OPE}
\label{sec:OPE.to.anom.dim.derivation}

Instead of the above derivation, let us simply define the anomalous
dimensions of $C$ and $A$ by derivatives of the logarithms of $C$ and
$A$ with respect to $\ln\mu$:
\begin{align}
  \gamma_C(Q,\mu,\alpha_s)
  & = - \frac{ d\ln C }{ d\ln\mu },
\\
  \gamma_A(P,H,m,\mu,\alpha_s)
  & = - \frac{ d\ln A }{ d\ln\mu }.
\end{align}
Since we do not use the connection to the renormalization factor
$Z_A$, we are no longer guaranteed that the anomalous dimensions are
independent of $Q$, $P$, $H$, $m$, and $\mu$: Whatever arguments are
needed for $C$ and $A$ are also needed for $\gamma_C$ and $\gamma_A$
respectively.  If this dependence actually existed, then the solution
(\ref{eq:RG.solution}) of the RG equation would not be very useful,
since large logarithms could appear in the calculation of $\gamma_A$ in the
exponent in the solution.  Moreover, the RG evolution could depend on
hadron flavor.

But the product $CA$ is RG invariant, from which it follows that the
sum of the anomalous dimensions is zero:
\begin{equation}
  \gamma_C(Q,\mu,\alpha_s) + \gamma_A(P,H,m,\mu,\alpha_s) = 0.
\end{equation}
Since $\gamma_A$ is independent of $Q$, differentiation with respect to $Q$
shows that $\partial \gamma_C/\partial Q = 0$, i.e., that $\gamma_C$ is independent of $Q$.
Similarly $\gamma_A$ is independent of $P$, $H$, and $m$, because $\gamma_C$ is.
Dimensional analysis then shows that neither anomalous dimension
depends on the explicit argument $\mu$.

In this derivation, the statement of the RG equation for $A$ is a
triviality: it is merely a definition.  The non-trivial result is that
$\gamma_A$ is a function only of $\alpha_s$.  From the point of view of
applications, it is this last property that is the critical one, since
it allows a perturbative calculation of $\gamma_A(\alpha_s(\mu))$ in Eq.\
(\ref{eq:RG.solution}) provided only that $\alpha_S(\mu)$ is large enough,
i.e., that $Q$ and $\mu_0$ are in what is often called the perturbative
region of mass scales for QCD.

In the first derivation, this same property followed from the
multiplicative renormalizability of the operator $\mathcal{O}$ and
from properties of \MSbar{} renormalization.  But in the second
derivation, renormalization properties are not used.  The contrast
between these two facts becomes less strange when one observes that
(a) the derivation of the OPE more-or-less determines which operator
is used, and (b) the requirement that the coefficient function
correspond to behavior that is purely leading-power in $Q$ requires
the use of a mass-independent renormalization scheme.  Furthermore the
proof of multiplicative renormalizability of the operator has a lot in
common with the proof of the OPE.

\section{Statement of TMD factorization, etc}
\label{sec:statement}

In this section, I will give all the statements of TMD factorization
and of the associated evolution equations together with the other
results that are needed for full phenomenological application.  They
will be presented in the generalized CSS form that was derived (to all
orders) in \cite{Collins:2011qcdbook}.  The forms given by BN
\cite{Becher:2010tm} and by EIS \cite{GarciaEchevarria:2011rb} are
similar (except that BN do not define individually finite TMD pdfs).
The justifications given by BN and EIS are, of course, different, but
that does not affect the structural properties stated here.

\subsection{TMD factorization in CS notation}

The Drell-Yan process is production of a high-mass lepton pair via a
virtual electroweak boson of momentum $q$ in a high-energy collision
of two hadrons $A$ and $B$.  The factorization formula for the cross
section has the form
\begin{multline}
\label{eq:TMD.fact}
  \frac{d\sigma}{d^4q}
  =
  \sum_{ij} H_{ij}\big(Q^2/\mu^2,\alpha_s(\mu)\bigr)
      \int d^2\3{b}_T e^{i\3{q}_T\cdot\3{b}_T}
\times \\ \times 
      \tilde{f}_{i/A}(x_A,\3{b}_T; \zeta_A,\mu)
      \tilde{f}_{j/B}(x_B,\3{b}_T; \zeta_B,\mu)
  + Y,
\end{multline}
and is valid up to corrections suppressed by a power of $Q$.  Here,
$H_{ij}$ is the hard-scattering associated with the process at the
partonic level, which is just a quark-antiquark annihilation to the
lepton pair (or a gluon-gluon hard-scattering for Higgs production).
The sum is over all relevant parton flavors.  The quantities
$\tilde{f}$ are the TMD pdfs Fourier transformed into transverse
coordinate space, while $x_A$ and $x_B$ are longitudinal momentum
fractions, $x_A=Qe^{y_q}/\sqrt{2}P_A^+$, $x_B=Qe^{-y_q}/\sqrt{2}P_B^-$,
where standard light-front coordinates are used,\footnote{I use the
  convention that $V^{\pm}=(V^0\pm V^3)/\sqrt2$.  Note that up to
  corrections suppressed by a power of $q_T/Q$, $x_A=q^+/P_A^+$,
  $x_B=q^-/P_B^-$.  Using this last two equations to define $x_A$ and
  $x_B$ would keep the structure of TMD factorization, but would
  require a change in the calculation used in the $Y$ term.  }
$y_q=\frac12\ln(q^+/q^-)$ is the rapidity of the lepton pair, and $P_A$
and $P_B$ are the 4-momenta of the beam hadrons.  The quantity $\mu$ is
the renormalization scale, which we will assume to be in the \MSbar{}
scheme.  The cross section is RG invariant, i.e., it is independent of
the choice of $\mu$.

The parameters $\zeta_A$ and $\zeta_B$ are defined by
\begin{subequations}
\label{eq:zetaAB.values}
\begin{align}
  \zeta_A & = 2x_A^2(P_A^+)^2e^{-2y_n} = Q^2e^{2(y_q-y_n)} ,
\\
  \zeta_B & = 2x_B^2(P_B^-)^2e^{2y_n} = Q^2e^{2(y_n-y_q)} ,
\end{align}
\end{subequations}
where $y_n$ is an arbitrarily chosen quantity, which in the CS and
Collins derivations of factorization can be thought of as a rapidity
defining a separation between left-moving quanta from right-moving
quanta.  Technically, $y_n$ is the rapidity of a vector $n^\mu$ that
implements the separation, either by a choice of gauge-fixing or as
the direction of certain Wilson lines.

The first term on the right-hand side of Eq.\ (\ref{eq:TMD.fact}) is
the TMD factorization result. It gives a good approximation to
cross-section when $q_T \ll Q$.  The second term, $Y$, was proposed by
CS to provide a matching of TMD factorization with ordinary collinear
factorization for large $q_T$.  That is, it allows one to combine TMD
factorization with collinear factorization, so that Eq.\
(\ref{eq:TMD.fact}) is valid up to corrections suppressed by a power
of $Q$ for any $q_T$.  For the rest of this paper, the $Y$ term (or
some other matching method) will be ignored.  The angle of the lepton
pair has been integrated over, and the hadrons are assumed to be
unpolarized.  Simple generalizations of TMD factorization obeying the
same principles can be readily worked out to deal with angular
dependence and polarization (e.g., \cite{Aybat:2011ge}).

The hard scattering depends only on $Q$, $\mu$ and the strong coupling
$\alpha_s(\mu)$, and on the flavors of the annihilating partons.  If $\mu$ is
chosen to be of order $Q$, then perturbative calculations are
appropriate for it, with the lowest order just being the parton model
calculation for whatever version of the Drell-Yan process is being
used.  In contrast, the TMD pdfs contain all the contributions to the
cross section (at leading power in $Q$) from the non-perturbative
domain of QCD.  They depend, therefore, on quark masses and the
flavors of the partons and hadrons.  

\subsection{Evolution equations, and small $b_T$ expansion}

The TMD pdfs in Eq.\ (\ref{eq:TMD.fact}) depend on two auxiliary
parameters $\zeta$ and $\mu$.  To obtain useful predictions for experiments
at different energies, equations are needed to relate the TMD
functions at different values of the auxiliary parameters.  In
addition there is an equation that is a kind of generalized operator
product expansion (OPE) that expresses the TMD pdfs in terms of the
ordinary integrated pdfs when $b_T$ is small enough to be considered
to be in the region where perturbative calculations are appropriate in
QCD.

The CS equation is for the evolution with respect to $\zeta$ of a TMD pdf
for a parton of flavor $i$ in a hadron $A$.  In the form appropriate
to the definitions in Ref.\ \cite{Collins:2011qcdbook} it is
\begin{equation}
  \label{eq:TMD.pdf.CS}
  \frac{ \partial \ln \tilde{f}_{i/A}(x_A,b_T; \zeta_A, \mu) }
       { \partial \ln \sqrt{\zeta_A} }
  = 
  \tilde{K}(b_T;\mu),
\end{equation}
with an exactly similar equation for the TMD in hadron $B$, of course.
The quantity $\tilde{K}$ is a new entity, which I call the kernel of
the CS equation.  In Ref.\ \cite{Collins:2011qcdbook}, it was
constructed in terms of a vacuum matrix element of a certain
Wilson-loop operator.  It is independent of $x$ and $\zeta$, and of the
flavor of the parton and the hadron in the pdf.  But it does depend on
the color representation for the parton, i.e., there is one
$\tilde{K}$ for quarks and another for the gluon.

(It is worth observing that the TMD pdfs and $\tilde{K}$ depend also
on the coupling $\alpha_s(\mu)$ and the quark masses, as well as on the
parameters explicitly indicated.)

In the CS style of derivation, the CS equation is obtained by
differentiating with respect to the auxiliary rapidity parameter
$y_n$, because this can be exploited to make a relatively simple
derivation, using factorization properties and Ward identities.  This
appears to tie the CS equation to the use of a particular axial gauge,
in Ref.\ \cite{Collins:1981uw, Collins:1981uk}, or to the use of
certain non-light-like Wilson lines, in Ref.\
\cite{Collins:2011qcdbook}. But a physical interpretation is more
easily made if one considers holding $y_n$ fixed, e.g., at the
rapidity of the Drell-Yan pair.  Then the derivative with respect to
$\zeta$ becomes a derivative with respect to the physical momentum of the
beam hadron, actually with respect to its rapidity $y_{p_A}$:
\begin{multline}
  \label{eq:TMD.pdf.CS1}
  \frac{ \partial \ln \tilde{f}_{i/A}(x_A,b_T; \zeta_A, \mu) }
       { \partial \ln P_A^+ }
  = \frac{ \partial \ln \tilde{f}_{i/A}(x_A,b_T; \zeta_A, \mu) }
       { \partial y_{P_A} }
\\
  = 
  \tilde{K}(b_T;\mu).
\end{multline}

The kernel of the CS equation has an RG equation:
\begin{equation}
  \label{eq:TMD.K.RG}
  \frac{ d\tilde{K} }{ d\ln \mu }
  = -\gamma_K\xleft(g(\mu)\right).  
\end{equation}
The TMD pdfs obey the following RG equation:
\begin{equation}
  \label{eq:TMD.pdf.RG}
  \frac{ d \ln \tilde{f}_{i/A}(x_A,b_T; \zeta_A, \mu) }
       { d \ln \mu }
  = 
   \gamma_f\big( \alpha_s(\mu);\zeta_A/\mu^2 \bigr).
\end{equation}
In the \MSbar{} scheme, the anomalous dimensions $\gamma_K$ and $\gamma_f$ are
independent of quark masses.  

By observing that derivatives with respect to $\zeta$ and $\mu$ commute, it
follows \cite{Collins:1981uw, Collins:1981uk} that the dependence on
$\zeta$ of the anomalous dimension of the TMD pdf in Eq.\
(\ref{eq:TMD.pdf.RG}) is linear in $\ln\zeta$, and that the coefficient is
half $\gamma_K$:
\begin{equation}
\label{eq:gamma.f.zeta.dep}
    \gamma_f\big( g(\mu); \zeta/\mu^2 \bigr)
    = \gamma_f( g(\mu); 1 )
      - \frac12 \gamma_K(g(\mu)) \ln \frac{ \zeta }{ \mu^2 }.
\end{equation}

When $b_T$ is large enough to be considered non-perturbative, the TMD
pdfs and the CS kernel $\tilde{K}$ are non-perturbative, i.e., they
are quantities that at present must be obtained from fits to data and
not from perturbatively based calculations.

But at small $b_T$, $\tilde{K}$ can be evaluated perturbatively
provided that $\tilde{K}$ is evolved to where $\mu$ is of order
$1/b_T$ (so that the coupling is small and there are no large
logarithms of $b_T\mu$).  There is also an OPE for the TMD pdfs at
small $b_T$ in terms of ordinary pdfs $f_{j/H}(x;\mu)$:
\begin{multline}
  \label{eq:TMD.pdf.small.b}
  \tilde{f}_{i/H}(x,b_T;\zeta;\mu) 
=\\
  = \sum_j \int_{x-}^{1+} \frac{ d{\hat{x}} }{ \hat{x} }
       \,\tilde{C}_{i/j}\xleft( x/\hat{x},b_T;\zeta,\mu,g(\mu) \right)
       \, f_{j/H}(\hat{x};\mu),
\end{multline}
up to errors that are power suppressed in $b_Tm$.
Here the notations $x-$ and $1+$ indicate, as in
\cite{Collins:2011qcdbook}, that the integrals should be extended
slightly beyond the indicated limits to allow a treatment of
singularities in $\tilde{C}$ and $f$ as those of generalized functions
(i.e., distributions) in the mathematical sense.
The coefficient
function $\tilde{C}$ is perturbatively calculable at small $b_T$,
provided that the pdf is evolved to appropriate values of $\zeta$ and $\mu$.
CS and RG equations for $\tilde{C}$ follow from those for the TMD pdfs
and the DGLAP equation for the ordinary pdf.  The lowest order term in
$\tilde{C}$ corresponds to the statement that in the parton model, the
integral of the TMD pdf over all transverse momentum would equal the
collinear pdf.

\section{Properties to be proved}
\label{sec:to.prove}

Suppose we ignored the particular derivations of the CS and RG
equations, (\ref{eq:TMD.pdf.CS}), (\ref{eq:TMD.K.RG}), and
(\ref{eq:TMD.pdf.RG}).  Then we could just define the CS kernel
$\tilde{K}$ by Eq.\ (\ref{eq:TMD.pdf.CS}) and the anomalous dimensions
by Eqs.\ (\ref{eq:TMD.K.RG}) and (\ref{eq:TMD.pdf.RG}).  But that
would allow the CS kernel $\tilde{K}$ to depend on $x$, $\zeta$ and the
flavors of the parton and hadron.  It would allow the RG coefficients
$\gamma_f$ and $\gamma_K$ to depend on $x$, $b_T$, masses, and the hadron
flavor.  It would also not restrict the anomalous dimension $\gamma_f$ to
be linear in $\ln\zeta$, and $\gamma_K$ to be independent of $\zeta$.  The extra
dependencies, particularly of $\tilde{K}$ would remove much of the
predictive power of TMD factorization.  General properties of
renormalization do require the anomalous dimensions to be independent
of masses in the \MSbar{} scheme.

We therefore need to derive the follow properties:
\begin{enumerate}
\item $\tilde{K}$ is independent of $x$ and $\zeta$, and of parton and
  hadron flavor. 
\item $\gamma_K$ is independent of these same variables.
\item $\gamma_f$ is independent of $x$, of masses, and of hadron flavor. 
\item Its dependence on $\zeta$ is linear in $\ln\zeta$, as in Eq.\
  (\ref{eq:gamma.f.zeta.dep}) with the coefficient being $-\frac12\gamma_K$.
  Note that since $\gamma_f$ is independent of quark masses, the dependence
  on $\zeta$ and $\mu$ is only via the ratio $\zeta/\mu^2$, by dimensional
  analysis, and also $\gamma_K$ is independent of masses.
\end{enumerate}
The second property is a trivial consequence of the first, by the
definition (\ref{eq:TMD.K.RG}) of $\gamma_K$.

\section{Derivation of form of CS equation and RG kernel}

The TMD pdfs depend not only on parameters determined by the kinematic
variables of the process, but also on the auxiliary parameter $y_n$,
via the values of the $\zeta$ parameters in Eqs.\
(\ref{eq:zetaAB.values}).  But the cross section does not depend on
$y_n$.  Moreover, if one uses the form of TMD factorization derived in
\cite{Collins:2011qcdbook} or the similar factorization used in
\cite{Becher:2010tm} and \cite{GarciaEchevarria:2011rb}, the hard
factor is also independent of $y_n$; it depends only on $Q$, on the
renormalization scale $\mu$, and on the QCD coupling.  There is also no
separate soft factor in the factorization formula, unlike many of the
earlier formalisms, notably that of CSS \cite{Collins:1981uw,
  Collins:1981uk, Collins:1984kg}.  I now show how to use these
results to demonstrate the properties listed in Sec.\
\ref{sec:to.prove}.

The starting point of the derivation is not an absolutely necessary
property of a TMD factorization formalism.  For example, there can be
a soft factor.  Thus in some formalisms, e.g., that of Korchemsky
\cite{Korchemsky:1988hd} for the Sudakov form factor, it is the soft
factor that carries the $Q$-dependence that in Eq.\
(\ref{eq:TMD.fact}) is given by the $\zeta$ dependence of the TMD pdfs.
Moreover, in the formalism of Ji, Ma, and Yuan \cite{Ji:2004wu} there
appear several kinds of non-light-like Wilson lines, and not only does
that formalism have a soft factor but the hard scattering has
dependence on an extra auxiliary parameter $\rho$.

The new formalisms with a structure like that listed in Sec.\
\ref{sec:statement} are in some sense minimal (as in ``minimal
subtraction'').  The \emph{structure} applies equally to the EIS
formalism \cite{GarciaEchevarria:2011rb} and (modulo their lack of
finite individually defined TMD pdfs) to the BN formalism
\cite{Becher:2010tm}.  In both these last two cases, the TMD
factorization given by the authors is obtained by setting
$\zeta_A=\zeta_B=Q^2$ in Eq.\ (\ref{eq:TMD.fact}).  The existence of an
equivalent of a $y_n$ parameter is not obvious, but, at least in the
EIS case, can be considered a consequence of an implicit choice of
coordinate frame \cite{Collins:2012uy}.

Common themes of the new formalisms are that the Wilson lines (or
their equivalent) in the definitions of the various factors are made
light-like as far as possible (but with regulation at intermediate
stages), and that any soft factor is absorbed into the collinear
factors (i.e., the TMD pdfs).

\subsection{Derivation of CS equation}

To derive the form of the CS equation, I will follow the pattern used
in Sec.\ \ref{sec:OPE.to.anom.dim.derivation}, but now applied to
$y_n$ dependence instead of $\mu$ dependence.  A complication is that
there was one term in the OPE used in Sec.\
\ref{sec:OPE.to.anom.dim.derivation}, but there are generally multiple
terms in TMD factorization (\ref{eq:TMD.fact}), with its sum over
parton flavors.  The complication can be avoided by using the
following trick inspired by how flavor-separated pdfs can be obtained
from the structure functions of deep-inelastic scattering with charged
currents.

The derivation of TMD factorization applies equally when we use a
general current correlation function 
\begin{equation}
  \label{eq:Wmunu}
  W^{\mu\nu} = s \int d^4y e^{-iq\cdot z} 
     \langle AB; \mbox{in}| j^\mu(z) j^\nu(0) | AB; \mbox{in} \rangle.
\end{equation}
The initial state is the same high-energy state of two incoming
hadrons as before, but the current can be any that we choose to use,
not just one that carries the coupling to an electroweak boson.  In
the following, let us use the following chiral current giving a
transition between two particular quark flavors:
\begin{equation}
  \label{eq:j}
  j^\mu = \bar\psi_j \gamma^\mu (1-\gamma_5) \psi_i,
\end{equation}
with the quark flavor indices being unequal: $i \neq j$.  In this case,
TMD factorization involves only two terms in (\ref{eq:TMD.fact}),
which use the following combinations of TMD pdfs:
\begin{subequations}
\label{eq:pdf.combos}
\begin{align}
   \tilde{f}_{i/A}(x_A,\3{b}_T; \zeta_A,\mu)
   ~ \tilde{f}_{\bar{\jmath}/B}(x_B,\3{b}_T; \zeta_B,\mu),
\\
  \tilde{f}_{\bar{\jmath}/A}(x_A,\3{b}_T; \zeta_A,\mu)
  ~ \tilde{f}_{i/B}(x_B,\3{b}_T; \zeta_B,\mu).
\end{align}
\end{subequations}
That is, the $i$ quark may come from the $A$ hadron and the anti-$j$
quark from the $B$ hadron, or vice versa.

If the current were invariant (up to a sign change) under charge
conjugation invariance, then the two combinations of TMD pdfs in
(\ref{eq:pdf.combos}) would simply be added, with equal coefficients.
But this is not the case here.  Let us decompose the hadronic tensor
$W^{\mu\nu}$ into a sum of scalar structure functions multiplying
kinematic basis tensors.  With a standard choice of basis tensors, we
can find among the structure functions one for which factorization
uses the sum of the terms in (\ref{eq:pdf.combos}), and one for which
factorization uses the difference.  So we can construct a combination
of structure functions that uses just one term in its TMD
factorization formula:
\begin{widetext}
\begin{equation}
\label{eq:TMD.fact.simple}
  F
  =
  H\big(Q^2/\mu^2,\alpha_s(\mu)\bigr)
      \int d^2\3{b}_T e^{i\3{q}_T\cdot\3{b}_T}
      \tilde{f}_{i/A}(x_A,\3{b}_T; \zeta_A,\mu)
      ~ \tilde{f}_{\bar{\jmath}/B}(x_B,\3{b}_T; \zeta_B,\mu).
\end{equation}
(Alternatively, we could adjust the set of basis tensors so that this
is the coefficient of one of the new basis tensors.)  We work at low
transverse momentum, so we ignore the $Y$ term that would otherwise be
needed. If the pair of flavors $(i,j)$ is changed, the graphs for the
hard scattering change only by a permutation of quark flavor labels,
and have the same values, since masses are neglected in the hard
scattering.  Thus the value of $H$ is the same for all cases.

Let us define
\begin{subequations}
\label{eq:TMD.zeta}
\begin{align}
  \label{eq:TMD.zeta.A}
  L_{i/A}(x_A,\zeta_A,b_T;\mu)
  & = 
  \frac{ \partial \ln \tilde{f}_{i/A}(x_A,b_T; \zeta_A, \mu) }
       { \partial \ln \sqrt{\zeta_A} }
  = 
  - \frac{ d \ln \tilde{f}_{i/A}(x_A,b_T; \zeta_A, \mu) }
         { d y_n },
\\
  \label{eq:TMD.zeta.B}
  L_{\bar{\jmath}/B}(x_B,\zeta_B,b_T;\mu)
  & = 
  \frac{ \partial \ln \tilde{f}_{\bar{\jmath}/B}(x_B,b_T; \zeta_B, \mu) }
       { \partial \ln \sqrt{\zeta_B} }
  = 
  \frac{ d \ln \tilde{f}_{\bar{\jmath}/B}(x_B,b_T; \zeta_B, \mu) }
       { d y_n },
\end{align}  
\end{subequations}
\end{widetext}
where I have used no assumptions or knowledge about the dependence of
the left-hand sides on the $x$ and $\zeta$ parameters, and on the parton
and hadron flavors.  (There is also dependence the QCD coupling and
possibly on masses that is not indicated explicitly.)

Now neither the structure function $F$ nor the hard scattering factor
$H$ in Eq.\ (\ref{eq:TMD.fact.simple}) depends on $y_n$.  So
differentiating Eq.\ (\ref{eq:TMD.fact.simple}) with respect to $y_n$ gives
\begin{equation}
\label{eq:diff}
  0 = -L_{i/A}(x_A,\zeta_A,b_T;\mu) + L_{\bar{\jmath}/B}(x_B,\zeta_B,b_T;\mu) ,
\end{equation}
i.e., equality of the two kernels.\footnote{The obvious derivation of
  this equality assumes that the TMD pdfs are always non-zero.
  However, it is conceivable that they have nodes, i.e., zeros on a
  submanifold in the space of their arguments.  In that case
  $L_{i/A}=L_{\bar{\jmath}/B}$ is established by analytic continuation from
  where the TMD pdfs are nonzero. }

Next we observe that one $L$ depends on $x_A$ and $\zeta_A$, but not on
$x_B$ and $\zeta_B$, while the reverse is true for other $L$.  Since these
quantities can be varied independently,\footnote{The four quantities
  $x_A$, $\zeta_A$, $x_B$, and $\zeta_B$ can be changed independently by
  varying the following four quantities $s=(P_A+P_B)^2$, $Q$, the
  center-of-mass rapidity of the lepton pair, and $y_n$.} it follows
each $L$ is independent of its $x$ and $\zeta$ arguments.  Similarly, the
hadron flavors can be varied independently, so that the $L$ functions
are independent of hadron flavor.  Finally, by taking all possible
values of $i$ and $j$, we find that the $L$ functions for different
quark flavors are also equal, i.e., $L$ does not depend on quark
flavor.  So we can simply rename $L$ to $\tilde{K}$ to give the kernel
in Eq.\ (\ref{eq:TMD.pdf.CS}).

Notice that the argument would apply in hypothesized extensions of QCD
(e.g., supersymmetric extensions) to include for example color-triplet
scalars, and color-octet quarks and scalars.  In each of these cases,
we can construct gauge-invariant operators that couple one of the new
fields to one of the old fields, and it then follows that the $L$
function for a new field is the same as $\tilde{K}$ for an old field
with the same color.  Of course, fields with new color
representations, like color-sextet quarks, are not covered by this
argument; each color representation needs a new $\tilde{K}$ function.

The result of this section is that all the TMD pdfs obey
\begin{equation}
  \label{eq:TMD.pdf.CS.copy}
  \frac{ \partial \ln \tilde{f}_{i/A}(x,b_T; \zeta, \mu) }
       { \partial \ln \sqrt{\zeta} }
  = 
  \tilde{K}(b_T;\mu),
\end{equation}
where $\tilde{K}$ depends only on $b_T$, on the parameters of QCD,
including masses, and on the color representation for parton $i$.

The RG equation for $\tilde{K}$ will be derived in Sec.\
\ref{sec:K.RG}. 

One way of summarizing the main issue for the above results is that,
as regards kinematic variables, each TMD pdf depends, not only on the
standard kinematic variables $x$ and $b_T$, but separately on $P_A^+$
for $\tilde{f}_{i/A}$ and $P_B^-$ for $\tilde{f}_{j/B}$.  But by
Lorentz invariance, the cross section can only depend on the product
$P_A^+P_B^-$.  From the form of factorization used in this paper, it
follows that the product $\tilde{f}_{i/A}\tilde{f}_{j/B}$ depends on
the product $P_A^+P_B^-$, but not on $P_A^+$ and $P_B^-$ separately.
The proof shows that the only way that this can be accomplished is to
have a power-law dependence:
\begin{equation}
\label{eq:TMD.pdf.P.dep}
  \tilde{f}_{i/A} \propto (P_A^+)^{\tilde{K}},
\qquad
  \tilde{f}_{j/B} \propto (P_B^-)^{\tilde{K}},
\end{equation}
where the power may depend on $b_T$, but is independent of the flavors
of the partons and hadrons and is independent of other kinematic
variables.  Note that $P_A^+P_B^-=Q^2/(2x_Ax_B)$ up to mass
corrections.  

Now Eq.\ (\ref{eq:TMD.pdf.P.dep}) has a non-boost-invariant form.  But
the TMD pdfs also depend on the direction of the vector $n$ specifying
the non-boost-invariant operators in their definition.  So the Lorentz
invariance of QCD implies that the dependence on hadron momenta can be
put in the Lorentz-covariant form
\begin{equation}
\label{eq:TMD.pdf.P.dep2}
  \tilde{f}_{i/A} \propto \left( P_A^+ |n^-/n^+| \right)^{\tilde{K}},
\qquad
  \tilde{f}_{j/B} \propto \left( P_B^- |n^+/n^-| \right)^{\tilde{K}} .
\end{equation}

\subsection{Derivation of RG equation of TMD pdf}

Let us now define the anomalous dimension of a TMD pdf in hadron $A$
by Eq.\ (\ref{eq:TMD.pdf.RG}) and the exactly similar equation for the
TMD pdf in hadron $B$.  However, with this equation treated as the
definition, the anomalous dimension of a TMD pdf could depend on the
flavor of its beam hadron and its quark, on the value of $x$, and on
masses: e.g., $\gamma_{i/A}\big( x_A, \alpha_s(\mu);\zeta_A,\mu^2,m(\mu) \bigr)$.  We will
generalize the argument used for the elementary OPE in Sec.\
\ref{sec:OPE.to.anom.dim}.  It will show that $\gamma_{i/A}$ is independent
of the quark and hadron flavor, and that it is independent of $x$ and
of quark masses.  Mass independence of an anomalous dimension is also
a consequence of the infra-red safety of renormalization in a
mass-independent scheme, and infra-red safety is a prerequisite for
reliable perturbative calculations.  However, unlike the previous
case, dependence on $\zeta$ and $Q$ remains and needs to be analyzed.

As with our derivation of the CS equation, we use a structure which
has a single term in its TMD factorization,
(\ref{eq:TMD.fact.simple}).  Then, from the RG invariance of the cross
section, it follows that the sum of the anomalous dimensions of the
two TMD pdfs and of the hard scattering is zero:
\begin{equation}
\label{eq:sum.gam}
   \gamma_{i/A}\big( x_A, \zeta_A \bigr)
   + \gamma_{\bar{\jmath}/B}\big( x_B, \zeta_B \bigr)
   + \gamma_H\big( Q^2 \bigr)
   = 0,
\end{equation}
where $\gamma_H = dH/d\ln\mu$, and the arguments have been omitted for all
but the dependence on kinematic- and flavor-related variables.  We can
change $x_A$ and $x_B$ independently of the other variables, so
differentiating with respect to $x_A$ and $x_B$ shows that each
anomalous dimension of the TMD pdfs is independent of its $x$
variable.  Similarly, we can change the hadron flavors independently,
so that $\gamma_{i/A}$ is independent of $A$.

We have already seen that $H$ is independent of which values of $i$
and $j$ are used (provided that $i \neq j$).  So considering all the
possible values of $i$ and $j$, shows that the anomalous dimensions
are equal for all flavors of quark and for all flavors of antiquark.
The charge-conjugation invariance of QCD shows equality between the
anomalous dimensions for quarks and antiquarks.  Hence all the
anomalous dimensions of the quark and antiquark TMD pdfs are equal: 
$\gamma_{i/A} = \gamma_f$, with our previous notation. 

To show independence of $\gamma_{i/A}$ of masses, we set $\zeta_A=\zeta_B=Q^2$, and
use flavor independence in Eq.\ (\ref{eq:sum.gam}).  This gives
\begin{equation}
   \gamma_f\big( \alpha_s(\mu) ; Q^2, \mu^2, m(\mu) \bigr)
   = -\frac12 \gamma_H\big( \alpha_s(\mu), Q^2/\mu^2 \bigr).
\end{equation}
Since $H$ is independent of masses, it follows that $\gamma_f$ is.  Thus in
all cases, the functional dependence is $\gamma_{i/A} =
\gamma_f(\alpha_s(\mu),\zeta_A/\mu^2)$, as already asserted in Sec.\
\ref{sec:statement}. 

Unlike the case of the kernel of the CS equation, the proof of
flavor-independence only works within the set of color-triplet
spin-half quarks, since the hard scattering would be different in the
case of hypothesized scalar quarks, for example.  Then the anomalous
dimension of the hard scattering could be different, which would
require a changed value of $\gamma_{i/A}$ for the scalar quark.

To analyze the $\zeta$ dependence of $\gamma_f$ and the $Q$ dependence of
$\gamma_H$, we rewrite Eq.\ (\ref{eq:sum.gam}) with all the allowed
functional dependence exhibited:
\begin{multline}
  \gamma_f\big( \alpha_s(\mu);Q^2e^{2(y_q-y_n)}/\mu^2 \bigr)
  + \gamma_f\big( \alpha_s(\mu);Q^2e^{2(y_n-y_q)}/\mu^2 \bigr)
\\
  = - \gamma_H\big( \alpha_s(\mu), Q^2/\mu^2 \bigr) .
\end{multline}
The right-hand-side does not depend on $y_n$, because $H$ does not.
Therefore differentiating the twice with respect to $y_n$ and setting
$y_n=y_q$ gives
\begin{equation}
  \frac{ \partial^2\gamma_f\big( \alpha_s(\mu);\zeta/\mu^2 \bigr) }{ \partial(\ln\zeta)^2 } = 0
\end{equation}
Hence the dependence of $\gamma_f$ on $\ln\zeta$ is linear.

\subsection{RG for kernel of CS equation}
\label{sec:K.RG}

In this section, I relate the $Q$ and $\zeta$ dependence of the anomalous
dimensions $\gamma_f$ and $\gamma_H$ to the RG properties of $\tilde{K}$, and
derive the RG equation for $\tilde{K}$.  This involves an
energy-independent anomalous dimension $\gamma_K$.

Now derivatives with respect to different variables commute:
\begin{multline}
  \frac{ d }
       { d \ln \mu }
  \frac{ \partial }
       { \partial \ln \sqrt{\zeta} }
  \ln \tilde{f}_{i/A}(x,b_T; \zeta, \mu)
\\
 =
  \frac{ \partial }
       { \partial \ln \sqrt{\zeta} }
  \frac{ d }
       { d \ln \mu }
  \ln \tilde{f}_{i/A}(x,b_T; \zeta, \mu).
\end{multline}
From the RG and CS equations for the TMD pdfs we then get
\begin{equation}
  \frac{ d \tilde{K}(b_T;\mu)}
       { d \ln \mu }
 =
  \frac{ \partial \gamma_f\big( \alpha_s(\mu);\zeta/\mu^2 \bigr) }
       { \partial \ln \sqrt{\zeta} }.
\end{equation}
The left-hand-side is independent of $\zeta$.  The right-hand side is
independent of masses, and therefore the left-hand-side is a function
just of $\alpha_s(\mu)$.  We therefore have the form of the RG
equation for $\tilde{K}$, Eq.\ (\ref{eq:TMD.K.RG}).  The $\zeta$
dependence of the anomalous dimension of the TMD pdf immediately
follows, as given in Eq.\ (\ref{eq:gamma.f.zeta.dep}).  Then for the
anomalous dimension of the hard scattering we get
\begin{equation}
\label{eq:gamma.H.zeta.dep}
    \gamma_H\big( g(\mu), Q^2/\mu^2 \bigr)
    = -2 \gamma_f( g(\mu); 1 )
      + \gamma_K(g(\mu)) \ln \frac{ Q^2 }{ \mu^2 }.
\end{equation}

\section{Conclusions}

In this paper I have shown that the CS and RG equations for TMD pdfs
follow from structural properties of TMD factorization.  The primary
assumptions are that the formulation of TMD factorization uses only
the pure leading power of $Q$ and that no soft factor appears in the
factorization formula (unlike the case for the formulations in
\cite{Collins:1981uw, Collins:1981uk, Collins:1984kg,
  Korchemsky:1988hd, Ji:2004wu}).  In the new derivation, no direct
use is made of properties of the operators whose matrix elements
define the TMD pdfs.  Of course, which operators are used is strongly
constrained by the requirement that the resulting TMD pdfs can be used
in a correct and useful TMD factorization property obeying the listed
assumptions.  The critical property underlying the proof is that the
product of TMD pdfs in Eq.\ (\ref{eq:TMD.fact}) is boost invariant
even though the individual TMD pdfs are not: See Eqs.\
(\ref{eq:TMD.pdf.P.dep}) and (\ref{eq:TMD.pdf.P.dep2}) and the
associated comments.

As compared with the analogous and simpler case of the operators that
are used in the OPE, the TMD case is notable for anomalous dimensions
that depend on $Q$ (either directly or through the $\zeta_A$ and $\zeta_B$
parameters).  This property is strange given the normal situation that
anomalous dimensions are independent of kinematic variables.

The $Q$-dependence of the anomalous dimensions is intimately tied to
the fact that perturbative calculations of processes for which TMD
factorization applies have two ``Sudakov'' logarithms of $Q$ per loop
instead of the one logarithm that occurs in situations to which the
standard OPE applies.  This can be seen by expanding the hard
scattering to first order in $\alpha_s$.  When the anomalous dimension is
$Q$-independent, consistency with the RG equation (\ref{eq:RG.C})
implies that there is at most one logarithm of $Q$ in the one-loop
calculation.

Since one-loop vertex graphs in gauge theories have a double logarithm
of $Q$, something more general is needed with TMD factorization.  This
can be provided if the anomalous dimension of the TMD pdf depends on
$Q$ (or $\zeta$).  This then entails that the TMD pdfs themselves depend
on $Q$ (or $\zeta$) unlike the case for the operator matrix element $A$ in
the OPE (and its generalizations to collinear factorization).  If a
different definition of TMD pdfs were to be constructed such that they
have no $Q$ dependence, then the formulation of TMD factorization must
be generalized from Eq.\ (\ref{eq:TMD.fact}), e.g., to put the
relevant $Q$ dependence in a soft factor, as in
\cite{Korchemsky:1988hd}.

Once one finds that the anomalous dimension is $Q$-dependent, there is
a danger that the exponent in the solution of the RG is no longer
usefully calculable. This is because of the presence of logarithms of
$Q/\mu$ in the perturbative expansion of the anomalous dimension, and
the typical phenomenon that the number of logarithms increases with
the number of loops.  This problem does not in fact occur because the
anomalous dimension of the TMD pdfs has at most one logarithm
independently of the order of perturbation theory.  This property,
Eq.\ (\ref{eq:gamma.f.zeta.dep}), was known in the earliest work by CS
\cite{Collins:1981uw, Collins:1981uk}, and in this paper it is also
derived from TMD factorization.

In all of the proofs in this paper, it was assumed that all quark
masses are light, so that they can be legitimately neglected (to
leading power in $m/Q$) in the hard scattering.  The independence on
masses of the hard scattering was important to the proof.  But in
reality there are heavy quarks in QCD, whose masses may be comparable
to or larger than $Q$.  Now the standard derivations of TMD
factorization assume that all quark masses are small compared with
$Q$.  Evidently, a useful project that needs to be tackled is to
generalize the results to the case that there are heavy quarks.


\section*{Acknowledgments}
I am grateful to M.~G.~Echevarria, A.~Idilbi, T.~Rogers, I.~Scimemi
and A.~Sch\"afer for useful discussions. The work was supported in
part by the U.S. Department of Energy under the grant number
DE-SC0008745.

\bibliography{jcc}

\end{document}